\documentclass{aa}
\usepackage{graphicx}
\usepackage{txfonts}
\begin{document}
   \title{Simultaneous Absolute Timing of the Crab Pulsar at Radio
   and Optical Wavelengths}

   \subtitle{}

   \author{T. Oosterbroek\inst{1}
          \and
          I. Cognard\inst{2}
          \and
          A. Golden\inst{3}
          \and
          P. Verhoeve\inst{1}
          \and
          D.D.E. Martin\inst{1}
          \and
		  C. Erd\inst{1}
		  \and
          R. Schulz\inst{4}
		  \and
		  J.A. St\"{u}we \inst{5}
		  \and 
		  A. Stankov \inst{1}
		  \and
		  T. Ho \inst{4}
          }

   \offprints{T. Oosterbroek}

   \institute{ Advanced Studies and Technology Preparation Division, Science Directorate of the European
Space Agency, ESTEC-ESA, P.O. Box 299, 2200 AG, Noordwijk, The Netherlands\\
              \email{toosterb@rssd.esa.int}
\and
Laboratoire de Physique et Chimie de l'Environnement, CNRS, 3A avenue
de la Recherche Scientifique, F-45071 Orleans, France
\and
Computational Astrophysics Laboratory, I.T. Building, National
University of Ireland, Galway, Ireland
         \and
             Research and Scientific Support Department of ESA, ESTEC,
P.O. Box 299, 2200 AG Noordwijk, The Netherlands
\and
Sterrewacht Leiden, Niels Bohrweg 2, 2333 CA Leiden, The Netherlands
             }

   \date{Received March 6, 2008; accepted June 23, 2008}


  \abstract
   {The Crab pulsar emits across a large part of the electromagnetic
	 spectrum. Determining the time delay between the emission at
	 different wavelengths will allow to better constrain the site and
	 mechanism of the emission.We have simultaneously observed the 
	 Crab Pulsar in the optical with S-Cam, an instrument based on 
	 Superconducting Tunneling Junctions (STJs) 
     with $\mu$s time resolution and at 2 GHz using the Nan\c cay
  radio telescope with an instrument doing coherent dedispersion
and able to record giant pulses data.}
   {We have studied the delay between the radio and optical
    pulse using simultaneously obtained data therefore reducing
  possible uncertainties present in previous observations.}
   {We determined the arrival times of the (mean) optical and radio
  pulse and compared them using the tempo2 software package.}
   {We present the most accurate value for the optical-radio lag of
  255$\pm$21~$\mu$s and suggest the likelihood of a spectral dependence
  to the excess optical emission asociated with giant radio pulses.}
   {}


   \maketitle
%

\section{Introduction}

  The Crab pulsar was the first pulsar detected as a consequence of
  its occasional bright pulses (now known as Giant Radio Pulses) at
  radio wavelengths (Staelin \& Reifenstein \cite{staelin:1968}; the
  first pulsar ever detected was PSR J1921+2153
  Hewish et al.\ \cite{hewish:1968}). Later on pulsations have been detected at
  optical wavelengths (Cocke et al.; \cite{cocke:1969}) and throughout
  the electromagnetic spectrum at X-rays and $\gamma$-rays. One
  feature which is constant over the whole spectrum is the pulsed
  emission, but the details differ substantially. E.g.\ the strength
  of the interpulse, the presence of a pre-cursor to the main pulse
  (at low radio frequencies) change with wavelength. However, the
  precise timing of the pulse over many orders of magnitude of energy
  imposes severe constraints on the emission regions (and
  mechanism). For example Romani \& Yadigaroglu (\cite{romani:1995})
  have suggested that the pre-cursor originates at the polar cap,
  while the pulse and interpulse originate in the outer gap in the
  magnetosphere, with higher energy pulses being generated at
  significantly greater heights. This should then be reflected in the
  timing of the pulses at different energies. Small differences in
  pulse alignment allow to study the emission regions at relatively
  small scales (about 10 km per 0.001 phase in case of the Crab
  pulsar), and precise timing of pulsar light curves throughout the
  electromagnetic spectrum is thus a powerful tool to constrain
  theories of the spatial distribution of various emission regions: a
  time delay most naturally implies that the emission regions differ
  in position.

In recent years, it has become clear that the main and secondary
pulses of the Crab Pulsar (PSR J0534+2200) are not aligned in time at
different wavelengths. X-rays are leading the radio pulse by reported
values of 344$\pm$40~$\mu$s (Rots et al.\ \cite{rots:2004}; RXTE data)
and 280$\pm$40~$\mu$s (Kuiper et al.\ \cite{kuiper:2003}; INTEGRAL
data) and $\gamma$-rays are leading the radio pulse by
241$\pm$29~$\mu$s (Kuiper et al.\ \cite{kuiper:2003}; EGRET data. The
uncertainty in this value does not include the EGRET absolute timing
uncertainty of better than 100~$\mu$s). At optical wavelengths, the
observations presented a less coherent picture. Sanwal (1999) has
reported a time shift of 140~$\mu$s (optical leading the radio). The
uncertainty in this value is 20~$\mu$s in the determination of the
optical peak and 75~$\mu$s in the radio ephemeris. Shearer et al.\
(\cite{shearer:2003}) have reported a lead of 100$\pm$20~$\mu$s for
simultaneous optical and radio observations of giant radio
pulses. Golden et al.\ (\cite{golden:2000}) have reported that the
optical pulse {\it trails} the radio pulse by
$\sim$80$\pm$60~$\mu$s. Romani et al.\ (\cite{romani:2001}) conclude
that the radio and optical peaks are coincident to better than
30~$\mu$s, but their error excludes the uncertainty of the radio
ephemeris (150~$\mu$s). Oosterbroek et al.\ (\cite{oosterbroek:2006},
hereafter O06) have studied this issue in detail and found an optical
lead of 273$\pm$100~$\mu$s. In this paper we will use simultaneous
optical and radio observations at 2 GHz to further improve the
accuracy of this value.

We also study the possible changes in the peak emission coincident
with Giant Radio Pulses (hereafter GP) as previously reported by
Shearer et al.\ (\cite{shearer:2003}).

\section{Observations}
\label{sect:obs}

\begin{table}
\caption{Log of optical and radio observations. All times are in
  UT. The first two optical observations were performed with a
  different instrumental setting resulting in different instrumental
  delays (see text).}
\label{tab:log}
\centering
\begin{tabular}{c c c c c c c c} 
\hline\hline
Observation & exposure (s) & Start time (yyyy-mm-dd hh:mm:ss)\\
\hline
0001 & 179 &2007-09-16 04:05:49\\
0002 & 900 &2007-09-16 04:09:05\\
0003 & 900 &2007-09-16 04:25:58\\
0004 & 900 &2007-09-16 04:41:03\\
0005 & 900 &2007-09-16 04:56:09\\
0006 & 900 &2007-09-16 05:11:14\\
0007 & 900 &2007-09-16 05:26:19\\
0008 & 900 &2007-09-16 05:41:25\\
0009 & 373 &2007-09-16 05:56:30\\
0013 & 1800 & 2007-09-17 04:10:47\\
0014 & 1800 & 2007-09-17 04:41:42\\
0015 & 1800 & 2007-09-17 05:11:48\\
0016 & 1100 & 2007-09-17 05:41:53\\
\hline
Radio 1 & 3420 &  2007-09-15 05:17:10\\
Radio 2 & 3930 &  2007-09-16 05:12:50\\
Radio 3 & 3960 &  2007-09-17 05:08:40\\
Radio 4 & 3890 &  2007-09-18 05:04:50\\
\hline\hline
\end{tabular}
\end{table}

Optical observations were obtained with S-Cam3 (Martin et al.\
\cite{martin:2006}) mounted on the ESA Optical Ground Station (OGS)
telescope on Tenerife on September 16 and 17 2007. For test
purposes the first two observations were obtained in the so-called
FAST mode of the instrument. This mode only differs in the digital
filter used to obtain the pulse shape of the photons. In S-Cam3 each
photon gives rise to a bi-polar signal in the detector electronics
chain. the zero-crossing of this signal is used to time tag the
photon. The instrumental delays in both modes were determined by
triggering an LED using a GPS and registering the assigned time of the
observed photons (see also O06). The instrumental delay of S-Cam3
amounts to 39.7$\pm$0.5~$\mu$s in the instrumental mode which was used
for all but the first two observations, and 14.0$\pm$0.5~$\mu$s for
the first two (i.e. ``FAST mode'') observations. Note that these
values are different from those reported in O06, since a physically
different detector array and different instrument modes were used.
Observations 0009 and 0016 have a shorter duration than the typical
exposure time for the night since they were interrupted because of
rising sky background (twilight). Some thin cirrus was present at the
end of the first night (starting at around 5:30 UT), while the seeing
was somewhat better during the first night compared to the second night.

Radio observations at 2 GHz were performed using the Nan\c cay radio
telescope on September 15-18 2007. The observations were done over a
bandwidth of 64 MHz (from 2016--2080 MHz) in 16 channels.  
Data were coherently dedispersed over 4 MHz bandwidth channels
(providing 250 ns resolution) using a transfer function in the complex
Fourier domain of the recorded voltages (Cognard \& Theureau
\cite{cognard:2006}, Demorest \cite{demorest:2007}).  Using a
polynomial representation for the expected evolution of the apparent
rotational phase of the pulsar, data were folded to produce a profile
every minute. A daily profile is built by integration of the 1-minute
profiles obtained over the typical 1-hour duration of an
observation. During data processing, the instrumentation is able to
detect and store shortlived events such as giant pulses known to occur
on this pulsar. Radio observations at this high frequency allow to
minimize the effect of uncertainties in the Dispersion Measure (DM).
For a log of the optical and radio observations we refer to table
\ref{tab:log}.

\section{Analysis and Results}

\subsection{Optical TOAs}

\begin{figure}
\centerline{
\includegraphics[width=8.5cm,angle=-0]{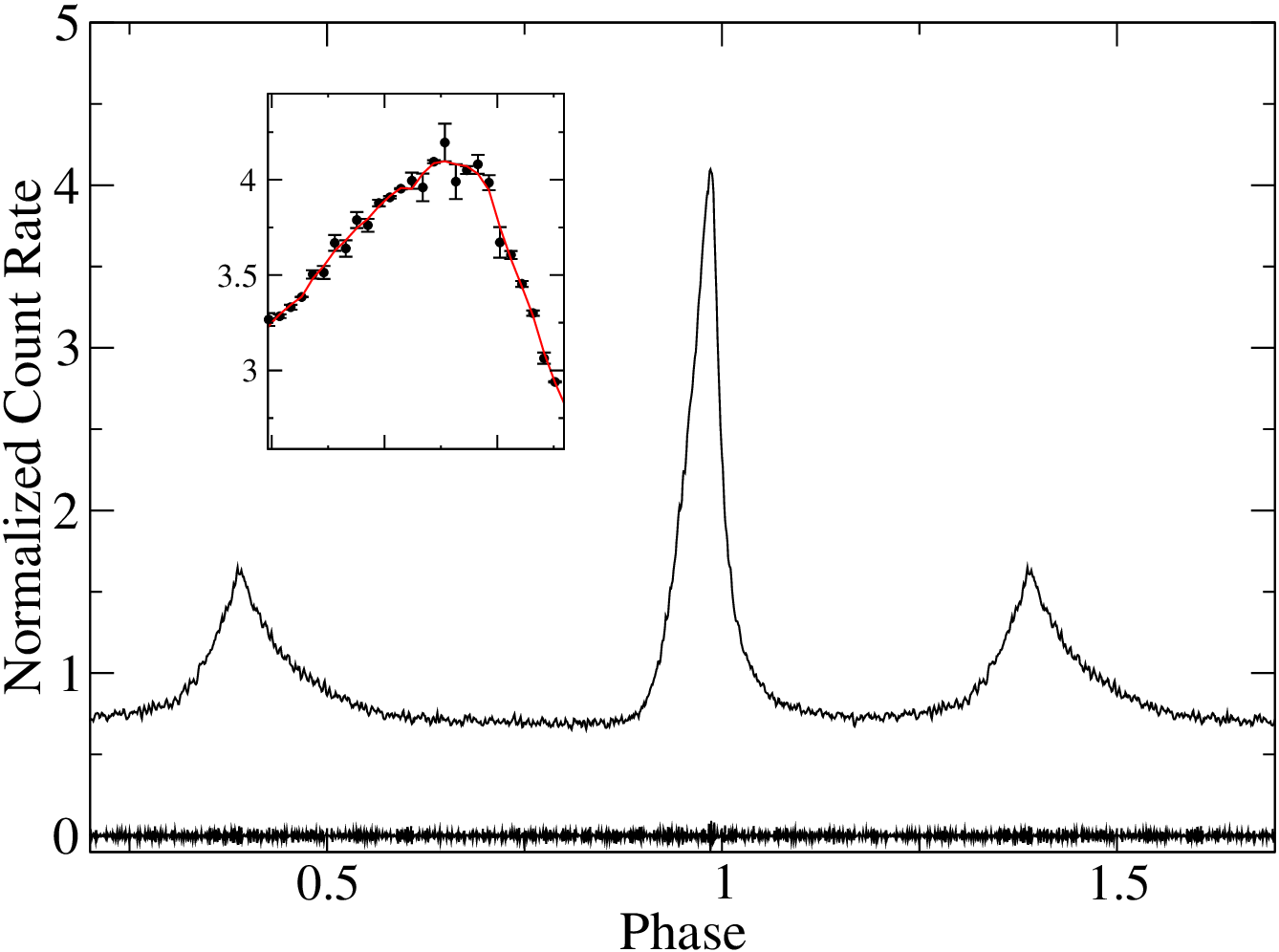}}
\caption[]{The pulse profile template used for obtaining the TOAs. The
  profile shown is the modestly smoothed profile (see text) which at
  this scale is indistinguishable from the unprocessed data. In the
  zoom-in the difference between the data points (circles with error
  bars) and the smoothed data (line) is shown for the peak where the
  differences are largest. At the bottom of the main plot the
  difference between the smoothed and non-smoothed data is shown.}
\label{fig:template}
\end{figure}

The analysis of the optical data consisted of searching for the best
{\it local} period by using a profile-folding technique for each S-Cam
observation. Since these observations are relatively short (up to
1800s) we did not take into account the period derivative nor did we
perform any barycentering on the time stamps. The data was then folded
on the best local period, using the midpoint of the observation as the
epoch. This resulted in 13 profiles, one for each observation. A
template profile was constructed. This template was created by applying a barycentric correction
to the time stamps of the photons and folding on phase (i.e.\
calculating the phase of every photon using an epoch and period and
summing to obtain a phase histogram) using the
best period determined by folding the data on a range
of trial periods and maximizing the $\chi^2$ with respect to the mean.
  All data were used, except the two first observation, since they
  are obtained in a different instrumental mode. A cross-correlation
  between every observation and the template profile was performed and
  the delay/lead was determined by fitting a Gaussian around the peak
  of the cross-correlation curve. The time of arrival of each average
  pulse profile was then assigned to the midpoint of each
  observation. These TOA's were corrected for the instrumental delay
  of S-Cam (14.0$\pm$0.5~$\mu$s for the first two observations,
  39.7$\pm$0.5~$\mu$s for the remainder, see
  Sect. \ref{sect:obs}). Using the cross-correlation technique the
  most significant contribution to the uncertainty in the TOA is the
  uncertainty in the determination of the peak of the template
  profile, which amounts to 18~$\mu$s. This uncertainty was
  determined by fitting the data points with a Gaussian in a 0.01 wide
  phase range around the peak. This phase range was determined
  iteratively and choosen symmetrically around the peak (see also
  O06). The uncertainty represents the 1 $\sigma$ uncertainty on the
  peak position ($\Delta \chi^2=1$). This uncertainty is larger than
  the uncertainty introduced by the cross-correlation since for the
  cross-correlation the whole profile is used. The individual
  uncertainties from the cross-correlation are an estimated
  1.4--3.3~$\mu$s. The value of 18~$\mu$s can be considered as a
  systematical error on all the indivual optical points, since they
  are all determined using the same template profile.

In order to test whether some of the scatter in arrival times was
caused by the presence of noise in the template profile we constructed
a template profile using all data (except the first two observations)
which was modestly smoothed by a Savitzky-Golay filter (see
Fig. \ref{fig:template}). The filter was constructed using 4 forward
and 4 backward data points and degree 4 (see Press et al.\
\cite{press:1992}). This almost did not affect the results: arrival
times did not change by more than 4.3~$\mu$s and the change was
typically less than 2~$\mu$s in a seemingly random way. This
  implies that the obtained time shift does not depend on some modest
  smoothing of the template pulse profile.

\subsection{Radio TOAs}

The radio time of arrivals (TOAs) were derived for each day of
observations yielding 4 TOAs. Each TOA was obtained by doing a
cross-correlation of each daily profile with a mean profile
obtained by integration over many months of observation.  This
cross-correlation gives a time and a uncertainty ranging from 3 to
8~$\mu$s. The relatively low signal-to-noise ratio (around 20)
for the daily profiles explains the limited accuracy of the TOA
determinations.
The reference point in the mean profile was chosen to be the
phase bin with highest radio flux. Given the signal-to-noise ratio and
the width of the phase bins, we estimate that this introduces an
uncertainty of 8~$\mu$s which affects the ensemble of radio TOA in an
identical way (and does not disappear when averaged over multiple
TOAs).



The determination of the radio TOAs included the Nan\c cay
instrumental delay. This delay was estimated from several observations
of the millisecond pulsar J1824-2452 done with the Green Bank and the
Nan\c cay radiotelescopes at the same epoch.  A careful analysis of
the mean profiles obtained with the two similar instrumentations
installed at Green Bank and Nan\c cay (both Astronomy Signal Processor
-ASP-, Demorest, \cite{demorest:2007}) give a relative offset of
7.0$\pm$0.4~$\mu$s. Knowing accurately the Green Bank instrumental
delay (which amounts to 15.74~$\mu$s; Demorest,
\cite{demorest:2008}), we can derive the Nan\c cay
instrumental delay as 8.7$\pm$0.4~$\mu$s.

\subsection{Giant radio pulses and DM determination}

\begin{figure}
\centerline{\includegraphics[width=8.5cm,angle=0]{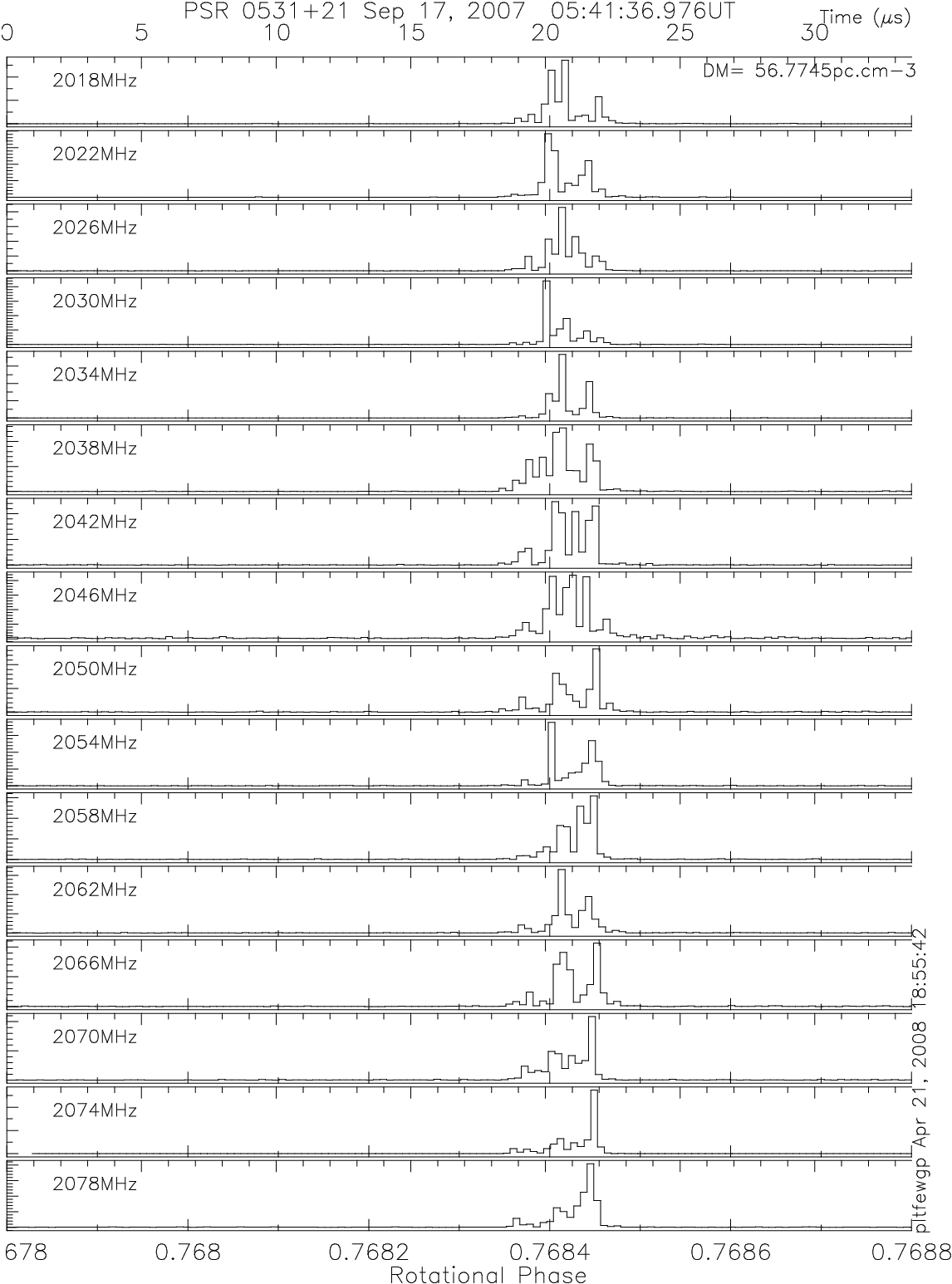}}
\caption[]{ Example of a giant pulse detected on Sept 17th with the
Nan\c cay radiotelescope. Total intensity is plotted for every 4 MHz
channels with a resolution of 250 ns. Duration of this giant pulse is
of the order of 2~$\mu$s.}
\label{fig:GP}
\end{figure}

Accurate determination of DM variations is usually done from TOAs
obtained at two different frequencies, as much separated as possible in
order to maximize the effect of the differential dispersion delay to
measure.  However, as shown by Ramachandran et
al.\ (\cite{ramachandran:2006}), highly frequency-dependent multipath
propagation in the turbulent interstellar medium produces huge
variations of the space (cigar shape like) within which the radio
waves senses the electron density.  TOAs obtained at very different
frequencies did not see exactly the same electron density and should
not be used to derive a highly reliable Dispersion Measure value.
Using frequencies as close as possible tends to remove this effect but
then TOAs determination must be much better to compensate.  As Giant
Pulses TOAs are characterized by very short bursts of radio emissions
suitable for a very good TOA determination, we should be able to use
them even on limited frequency span to determine unbiased DM
variations. For example, a differential delay between 2016 and
2080 MHz known at 0.5~$\mu$s gives roughly the same DM uncertainty as a
10~$\mu$s accuracy between 1400 and 2050 MHz.  

\begin{figure}
\centerline{
\includegraphics[width=6.5cm,angle=-90]{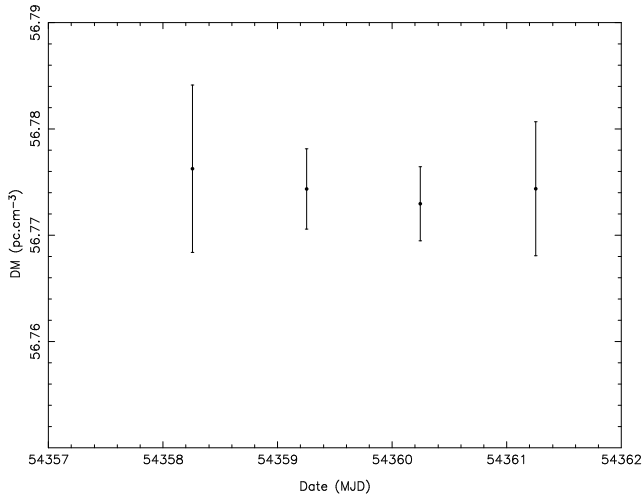}}
\caption[]{The DM value determined from the Nan\c cay GP data. Displayed
  are the nightly mean values.}
\label{fig:DMdetermination}
\end{figure}

We have determined the DM value from the Nan\c cay data alone in the
following way: each GP (an example is shown in Fig.\ \ref{fig:GP})
which was detected in more than 14 channels (among the 16 available,
using a 15 sigma threshold for each channel) was dedispersed with
different DM values and integrated in frequency to produce a 56 to
64 MHz total bandwidth profile. Based on the fact that a good DM value
will tend to properly align the individual GPs and thus maximize the
amplitude of the integrated profile, this maximum amplitude was kept
for each trial DM and a gaussian fit was done on those maximum
amplitudes to find the best DM. An average of the numerous DM obtained
for a given observation (one per single GP event) provides us with a
nightly DM value (see Fig.\ \ref{fig:DMdetermination}). Uncertainties
are estimated from the rms of the different DM values obtained each
night. The overall average value over the 69 DM values obtained over 4
nights is 56.7740 pc.cm$^{-3}$ and again an uncertainty of 0.005 is
estimated (from the rms over the 69 individual values).


\subsection{Optical-radio lag}

We determined an improved solution to the $\nu$ and
$\dot\nu$ starting from the Jodrell Bank ephemeris to the TOAs of the
mean pulse using the Nan\c cay radio data. While the fit to the 4
  data points is not very constraining, it establishes a reliable
  baseline against which the optical TOAs can be compared. 
The obtained values are $\nu =
29.755351174(9)$s$^{-1}$, and $\dot\nu = -3.727(6)
\times10^{-10}$s$^{-2}$ (for a fixed value of $\ddot\nu$ of $3.72
\times10^{-22}$s$^{-3}$). The dispersion measure was fixed at the
value reported above of 56.7740 pc cm$^{-3}$.

The OGS geocentric coordinates used in the reduction (within {\sc
  tempo2}) are (X,Y,Z) = (5390280.91, -1597890.59, 3007083.48). These
  coordinates were obtained from the reading of two different GPS
  units (see O06). The obtained optical TOAs were then used in
  the {\sc tempo2} analysis package (Hobbs et al.\ \cite{hobbs:2006}),
  together with the radio TOAs. The radio and optical data points
  were fit together allowing for an offset between the radio and
  optical. Since a comparison of arrival times is made at (almost) the
  same time, this procedure is not substantially affected by e.g.\
  uncertainties in the pulsar position (proper motion) or
  uncertainties in e.g.\ frequency and derivatives.

\begin{figure}
\centerline{
\includegraphics[width=7cm,angle=-90]{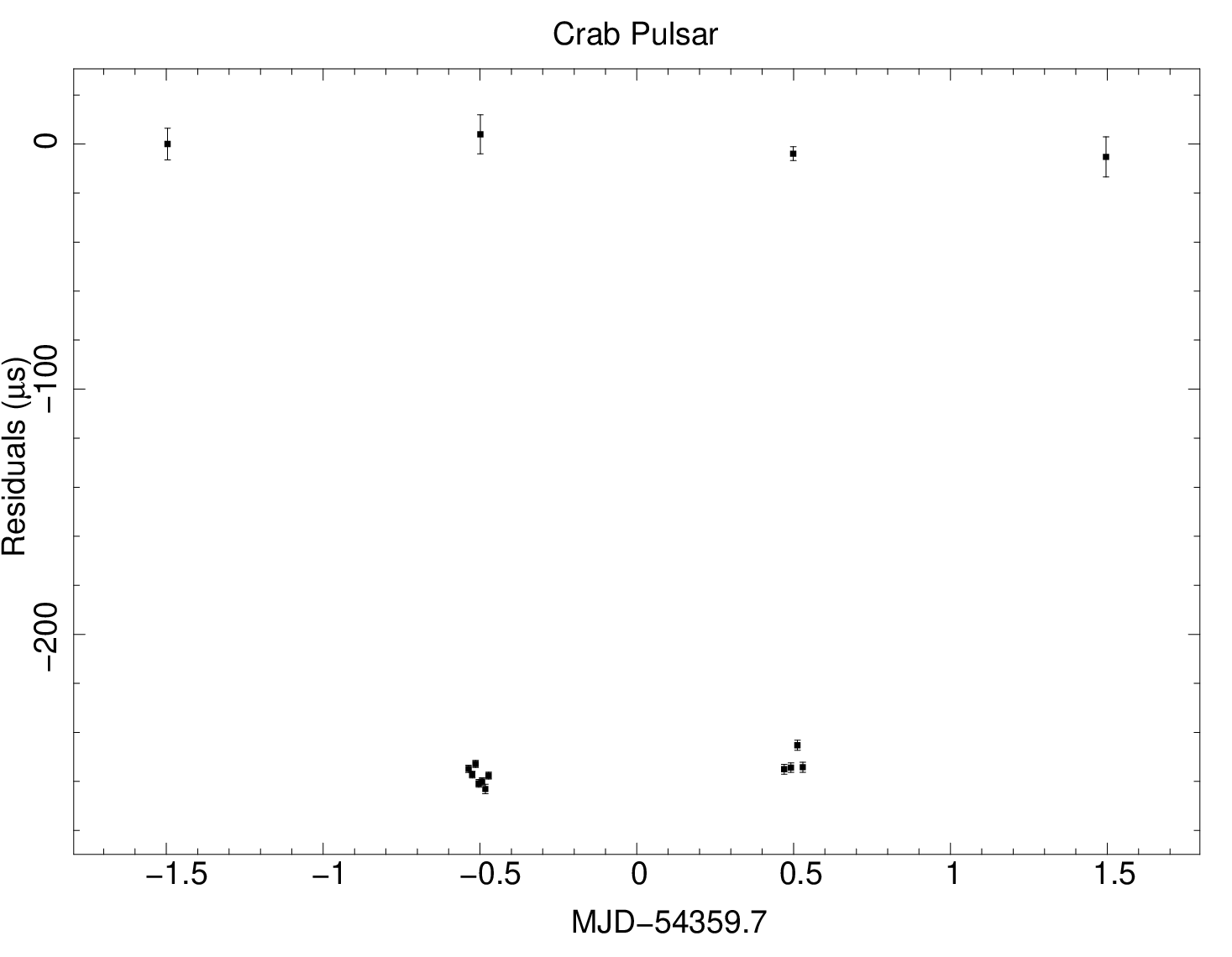}}
\centerline{
\includegraphics[width=7cm,angle=-90]{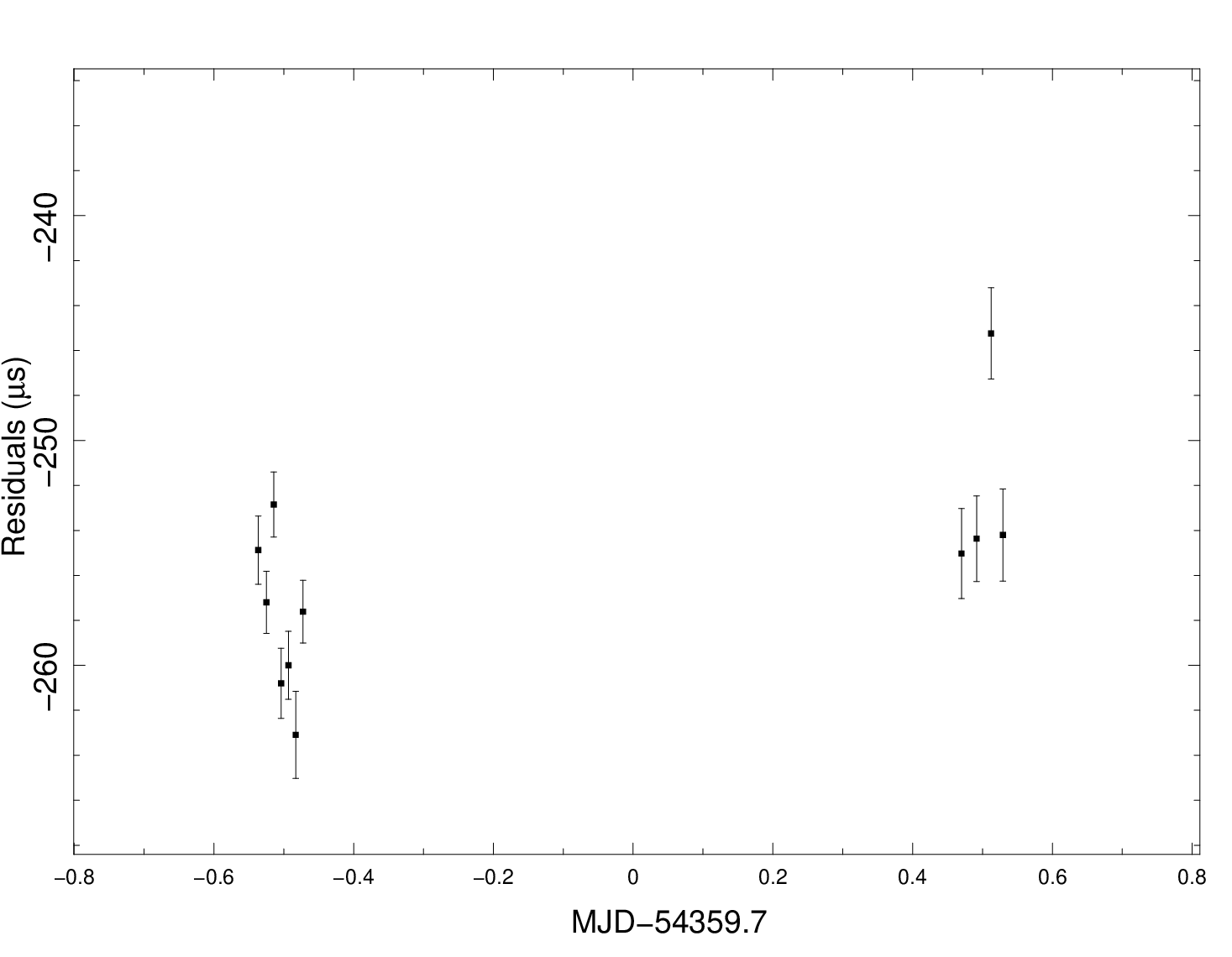}}
\caption[]{Comparison of radio and optical results. At the bottom a
  zoom-in on the optical data points only is presented. Plotted error
  bars represent 1 $\sigma$ uncertainties.}
\label{fig:results}
\end{figure}

\begin{figure}
\centerline{
\includegraphics[width=8.5cm,angle=-0]{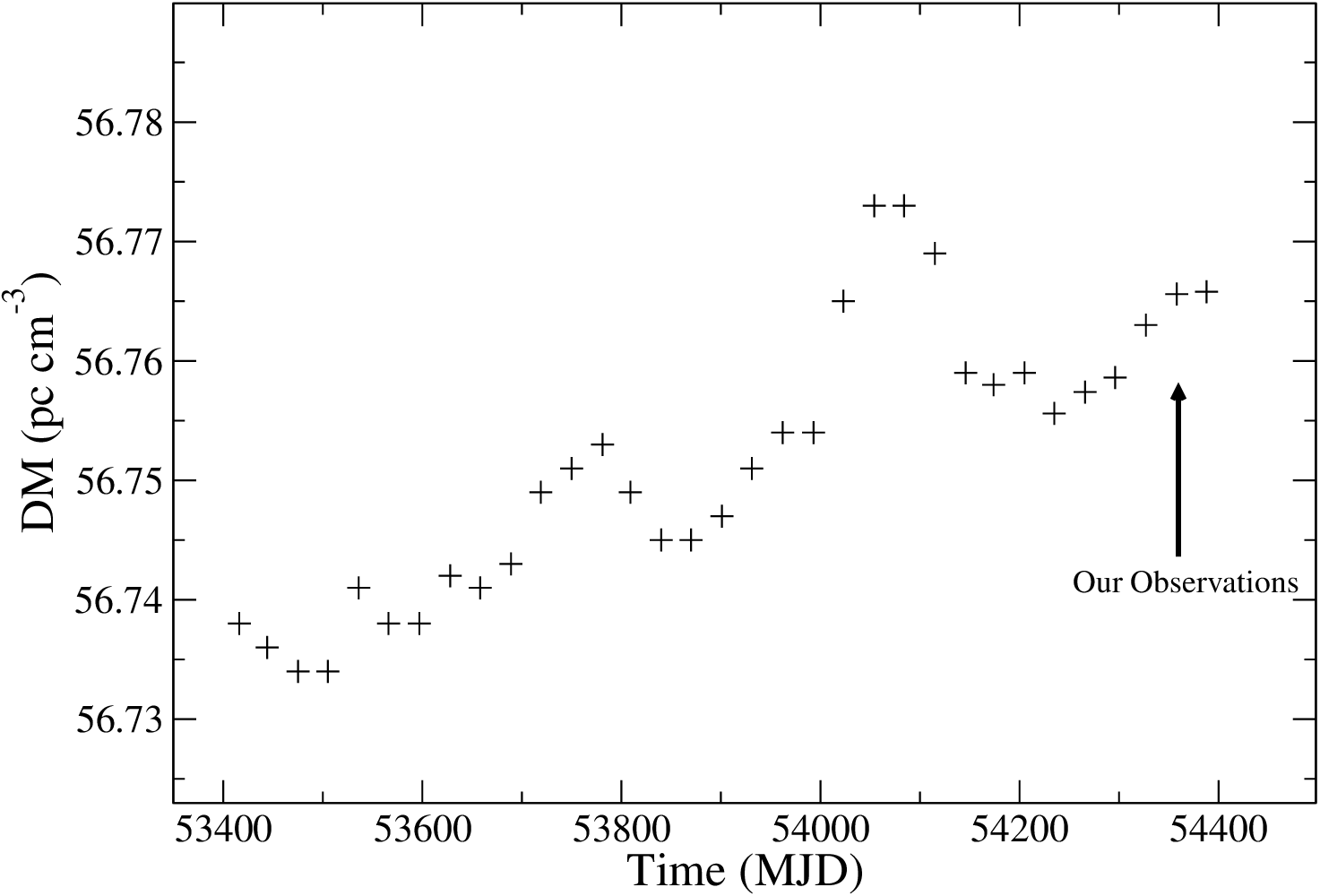}}
\caption[]{The recent (past 1000 days) history of the dispersion
  measure towards the Crab pulsar (data taken from Lyne et al.\
  \cite{lyne:2007}). The data points are monthly values. The time of
  our observations is indicated by an arrow. The difference
  between the Nan\c cay DM value and the value plotted here is less
  than 2 $\sigma$ (offset is 0.008 while uncertainty is 0.005 for both
  determinations). As the frequency spans used for these two
  derivations are not the same (610-1400 MHz for Jodrell and
  2018-2080 MHz for Nan\c cay), the difference is acceptable (see text).}

\label{fig:DM}
\end{figure}

In Fig. \ref{fig:results} the results of this analysis are displayed:
an offset of 255$\pm$4~$\mu$s between the radio peak and the optical
peak. The uncertainty in this value is purely the statistical
uncertainty in the best-fit value for an offset between the radio and
the optical measurements. This value does not take into account
uncertainties in the DM values, which could be of the same order. A
difference in the DM value of 0.005 pc cm$^{-3}$ (more than a typical
month-to-month variation, see Fig.\ \ref{fig:DM}) will result in a
time difference at 2 GHz of $\sim$6~$\mu$s. We therefore assume an
uncertainty of 6~$\mu$s as a result of uncertainties in the DM
value. This uncertainty in time is smaller than quoted by other
authors (e.g.\ Lyne et al.\ \cite{lyne:1993}), since our radio
observations have been obtained at a higher frequency which is
substantially less affected by timing delays due to scattering.

Adding all of the above uncertainties (18~$\mu$s from the optical
template, 8~$\mu$s from the radio template, 6~$\mu$s from DM
uncertainty, 4~$\mu$s from the fit to the offset between the radio and
optical TOAs) linearly we obtain a total uncertainty of
36~$\mu$s. This is likely to be an overestimate of the total
uncertainty since all individual deviations have to be in the same
direction to end up at this value. Adding all uncertainties in
quadrature yields a final uncertainty of 21~$\mu$s.

From Fig.\ \ref{fig:results} we conclude that the 
variation in the optical-radio delay is substantially larger than the
  individual uncertainties on the TOAs (which are between 1.4 and
  2.1~$\mu$s, 1$\sigma$). The average value of the delays obtained
  during the first and second night differ by 6~$\mu$s (only 1.4~$\mu$s
  when one excludes the deviant point in the second night). The
  observed scatter during the first night and the possible difference
  between the two nights, suggests that small variations (less than
  about 10~$\mu$s) exist. For this relative comparison the uncertainties
introduced by determining phase zero on both the radio and optical
templates (which each amount to 8~$\mu$s) should not be considered,
since they affect all data points in the same way. We consider it
unlikely that this night-to-night variation is solely caused by
variations in the DM, since a variation of $\sim$1 times that of typical
monthly variations is needed in 1 day. Variations in the peak position
on time scales of days and shorter are studied in more detail by
Karpov et al.\ (\cite{karpov:2007}), who find, during one
  observation, variations at the $\sim$150~$\mu$s-level.

\begin{figure}
\centerline{
\includegraphics[width=6cm,angle=-90]{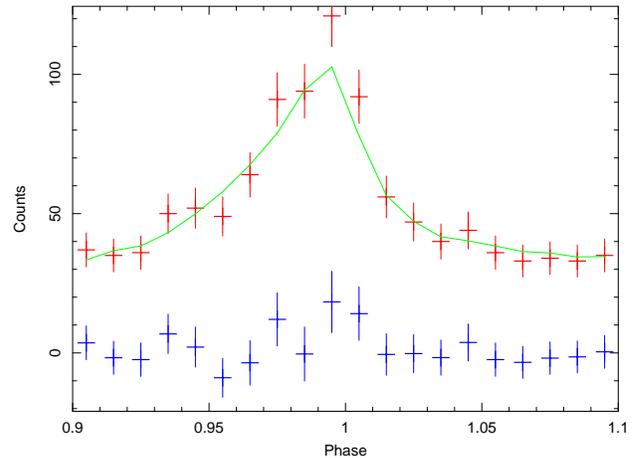}}
\caption[]{A comparison of the optical pulse profile obtained
  coincident with a Giant radio Pulse in red and a reference profile
  in green (see text). The difference between the two is also
  displayed.}
\label{fig:gp}
\end{figure}

\subsection{Giant radio pulses and optical}

In total 927 Giant
radio pulses were detected during the radio observations. The rate of
GPs was much higher during the second night where we have optical data
(on average 1 per 6.6 s) than the first night (1 per 29.5 s).

Following the method by Shearer et al.\ (\cite{shearer:2003}) we have
selected the optical data coincident with Giant Pulses (GPs) detected
in the radio observations. The barycentric arrival times of the GPs
were computed using {\sc tempo2}, while the event time tags were
corrected using our barycentering code. We checked that these
corrections are consistent to better than 15~$\mu$s and that therefore
this procedure provides the required accuracy. We find, in total, 603
GPs coincident with optical data. We made sure that these GPs occurred
near the main peak of the profile and selected data half a phase range
before and half a phase range after the main peak (in the radio). The
data was then folded to obtain a pulse profile coincident with radio
GPs. A reference profile was also constructed from data spanning 80
cycles around (but not including) this cycle. This was done since
changes in sky background and airmass force us to select data close to
the cycle coincident with the GP. Several different ranges were tried
and all gave similar results, with obviously better statistics as the
number was increased. We settled for 80 cycles, since this gives a
very good reference profile and yet it still only spans 2.7 s of
data. We ensured that no optical data was missing during these periods
(e.g. at the end of an observation). We present the obtained results
in Fig. \ref{fig:gp}. As can be seen from this figure no clear excess
is present. Depending on the choice of phase range this excess amounts
to about $\sim2.1\sigma$. We view this a confirmation of the result of
Shearer et al.\ (\cite{shearer:2003}) and shows that their dataset has
a much better statistical quality given the longer exposure times and
larger telescope aperture.

\subsection{Wavelength dependence of the delays}

\begin{figure}
\centerline{
\includegraphics[width=6cm,angle=-90]{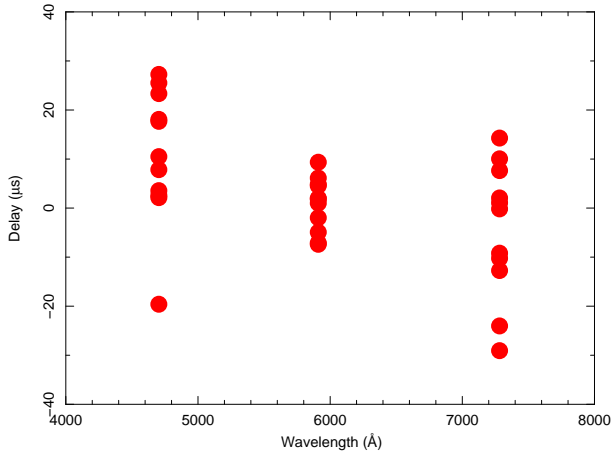}}
\caption[]{The delay between photons in different energy
  bands. Wavelengths indicate the center of the three bands (not the
  average energy of the photons).}
\label{fig:delays_color}
\end{figure}

In order to investigate whether we are able to detect a delay between
the pulse profiles obtained at different wavelengths we divided the
photons obtained during the observations described above in
three adjacent bands, each containing approximately the same number of
photons, using the intrinsic energy resolution of S-Cam. The
wavelength boundaries of the three bands were: 3920--5490 \AA,
5490--6330 \AA, 6330--8230 \AA. We then created the folded profiles
per observation and did the crosscorrelation with the smoothed profile
(which was obtained using all photons) as described above. The results
are displayed in Fig. \ref{fig:delays_color}. While the quality of the
data does not seem to justify drawing strong conclusions, a linear fit
to the datapoints reveals that the slope in the delay is
-5.9$\pm1.9$~$\mu$s/1000 \AA, which seems to indicate that the blue
photons arrive, averaged over phase, slightly earlier than the red
photons. A similar result is obtained when only the main peak part of
the pulse profile is used for this analysis. It is important to
realize that this is only a statement about the phase-averaged arrival
times of photons: the full width at half-maximum (FWHM) of the pulse
peak is known to decrease with decreasing wavelength (Percival et al.\
\cite{percival:1993}, Eikenberry et al.\ \cite{eikenberry:1997},
Sollerman et al.\ \cite{sollerman:2000}). This trend is also evident
in our data.
Together with the assymmetry in the profile the decrease in FWHM with
decreasing wavelength will lead to an average time shift. Close
inspection of our profiles in colour suggests that the profile rises
slightly earlier in the blue band, which could yield the result
present above. Furthermore the separation between the main and
secondary peak is smaller in the UV than in the optical (Percival et
al.\ \cite{percival:1993}). In essence this shows the limitation of
using the cross-correlation method: it is only possible to obtain a
proper time shift if the profile and template are identical in shape.
Any other method (e.g.\ fitting around the peak) will also be affected
by the assymetry of the profile and will give similar results.

\section{Discussion and Conclusions}

\begin{figure}
\centerline{
\includegraphics[width=8cm,angle=-0]{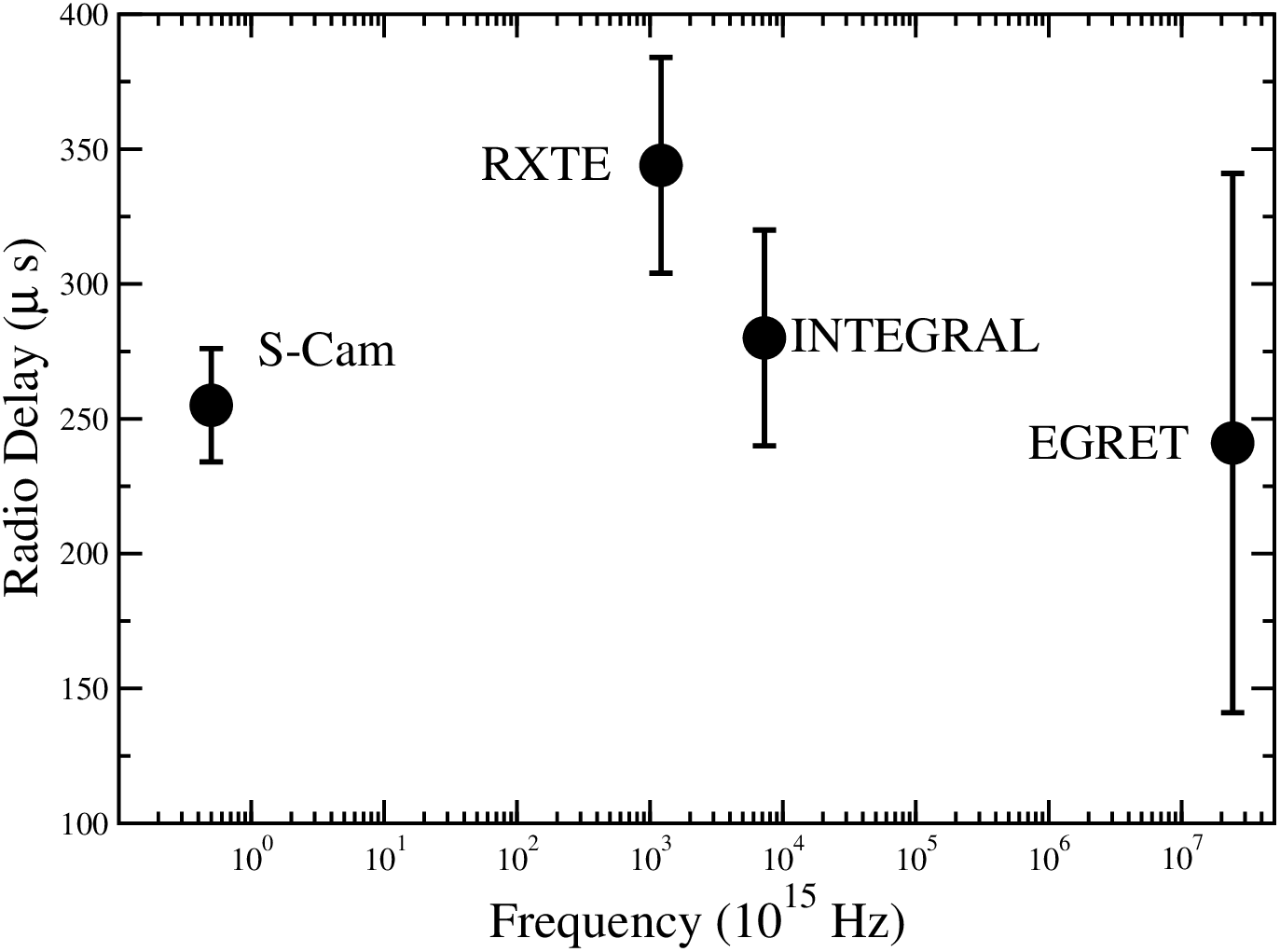}}
\caption[]{The radio delay with respect to other frequencies. The
  S-Cam data point is presented in this paper, the RXTE value is from
  Rots et al.\ (\cite{rots:2004}), the INTEGRAL and EGRET points are
  taken from Kuiper et al.\ (\cite{kuiper:2003}). We have added a
  100~$\mu$s error bar to the EGRET value, because of uncertainties in
  the EGRET absolute time calibration (see Kuiper et al.\
  \cite{kuiper:2003}).}
\label{fig:delays_em}
\end{figure}

We have performed simultaneous observations of the Crab pulsar in the
optical (3500--8200 \AA) and radio (2 GHz). The results provide the
current best estimate of the optical lead (with respect to the radio
peak) of 255$\pm$21~$\mu$s. The two major contributions to the
uncertainty are the determination (or definition) of the peak in the
optical, since the profile is asymmetric, and uncertainties in the DM
value which translate into significant uncertainty in the correction
of the radio data. Additionally a true jitter in the optical peak on
different time scales might take place with an amplitude of
5--10~$\mu$s. However, our value is an order of magnitude more precise
than determined by O06 and fully consistent with the value reported
there. The close match with their value (273$\pm$100~$\mu$s) must be
fortuitious given the large uncertainties on the values, and can be
considered a statistical coincidence. A comparison with previous
results can be be found in O06, however due to the increased precision
we can now conclude that the result of Shearer et al.\
(\cite{shearer:2003}), an optical lead of 100$\pm$20~$\mu$s, is not
consistent with our result. Together with the result of Golden et al.\
(\cite{golden:2000}), an optical {\it delay} of $\sim$80$\pm$60~$\mu$s,
which was already inconsistent with the results presented in O06, this
raises the possibility of a long-term change in the radio-optical
difference.

From a plot of the radio delay throughout the electromagnetic spectrum
(see Fig.\ \ref{fig:delays_em}) it appears that this delay is
approximately constant over more than 7 orders of magnitude. We
note that the non-optical results have been derived from a comparison
with, non-simultaneous Jodrell Bank observations, which results in
larger error bars. This near-constancy of the delay with respect to
the radio profile implies a similar emission mechanism and emission
location from the optical to $\gamma$-rays. This is indeed the case in
modern models by e.g.\ Takata \& Chang (\cite{takata:2007}).

Our results for the stability of the pulse profile (differences of up
to $\sim$10~$\mu$s on a time scale of a day and even smaller variations
on a time scale of less than a day; see Fig.\ \ref{fig:results}) are
not consistent with the results from Karpov et al.\
(\cite{karpov:2007}) who present variations of up to
$\sim$150~$\mu$s in a fraction of a day. However, this variation is
only present in a small part of their data set (December 2005 and
January 2006), while the remainder of their data (obtained in December
1994 and 1999, and November 2003) do not show substantial
variations. This suggest that the behaviour observed in
December-January 2005/2006 is an anomaly and that night-to-night
variations in the Crab optical peak positions amounts to at most
$\sim$10~$\mu$s.

Our value for the time delay between red and blue photons of
-5.9$\pm1.9$~$\mu$s/1000 \AA\ can be compared to the results of Sanwal
(\cite{sanwal:1999}) who found a delay between the R and U band of
-10$\pm$4~$\mu$s and Golden et al.\ (\cite{golden:2000}) who found that
the peaks in B and V are coincident within 10~$\mu$s. Taken together
these three measurements indicate a small delay between blue and red
photons. However changes in the FWHM of the pulse profile and the
assymmetry of the profile and possible differences in the separation
between the main and secondary peak make an interpretation complex.

Current models for high-energy pulsar emission (e.g.\ Dyks \&
Rudak \cite{dyks:2003}), describe the peaks as caustics: special
relativity effects (aberration of emission directions and time of
flight delays due to the finite speed of light) cause photons emitted
at different altitudes to pile up at the same pulse phase.  It is
difficult if not contrived to imagine how caustics originating from
the trailing edge extending along the full length of the open field
lines within the magnetosphere could produce phase bunching yielding a
common main peak cusp spanning 7 orders of magnitude in photon energy
from incoherent synchrotron emission, and less than 1\% in phase
difference from coherent radio emitting sources, spanning in this case
some 12 orders in photon energy.  As such arguments are usually
presented in the context of outer gap or slot gap model formalisms,
these new results argue strongly for a localised source for the
observed multiwavelength emission from the Crab pulsar.

That the radio and higher energy emission process are in someway
linked is not in doubt, following on from the optical-GP association
originally shown by Shearer et al. (\cite{shearer:2003}). Our
confirmation of an excess, albeit at low significance further
strengthens the connection between these energy bands. Variations in
the optical emission correlated with GPs have been argued to be a
consequence of local density enhancements to the plasma stream
(Shearer et al. \cite{shearer:2003}). Recent studies of GPs from the
Crab have indicated that these can and do originate as short-lived
($\sim$ ns), narrowband nanoshots, which have been argued to have
their origin in strong, highly localised plasma turbulence (Hankins \&
Eilek, \cite{hankins:2007}).  That there is no change in phase or
shape of the Crab's optical light curve morphology during a GP
associated event, just flux, requires an extended region of plasma
turbulence to produce the same phase-bunching effect following the
caustic model $ansatz$, which is clearly implausible.  The measured
time delay across the optical bands represents $\leq$ 3.0 $\times$
10$^{-4}$ of phase, with some evidence of a jitter in the optical peak
across several time scales occuring with a similar variation in terms
of phase. Again, these effects are not associated with any measurable
change in incident flux. Assuming a constant emission source,
localised geometric constraints represent the most likely cause of
such observational effects. Together these interpretations merit some
revision of the fundamental approaches to modeling the Crab pulsar's
emission processes, and there may be some use in trying to
geometrically constrain emission regions within the magnetosphere
first, and then address the local physical conditions that are
consistent with the data we have available to us - rather than the
other way round.

Finally the anticipated detection of excess optical emission
associated with GPs extrapolating from the work of Shearer et
al. (2003) was expected to be $<1\sigma$ prior to our observations, so
its marginal detection is somewhat surprising. We note that our
passband ($\sim$3500--$\sim$8000 \AA) is quite different from the
passband (6000--7500 \AA) of Shearer et al.\ (\cite{shearer:2003}) and
might result in different results depending on the spectrum of the
``excess'' optical emission.  If our observed excess had been
significant we could have constructed the optical spectrum obtained
when a GP occurs and compare it with the reference spectrum when no GP
is present. This is possible by using the intrinsic energy resolution
of S-Cam (each photon is tagged with a time stamp, but also its
energy, Martin et al.\ \cite{martin:2006}). Such an approach could
reveal the spectrum of the excess emission and consequently would
impose important further constraints on the emission mechanism. We
regard the acquisition of deeper coincident radio-optical data as
critically important to confirm the spectral dependence of such an
excess.

\begin{acknowledgements}
AG is supported by funding from Science Foundation Ireland
(SFI/RFP/PHYF553).  IC thanks Paul Demorest for fruitful interactions
on the instrumental delay determination and the support of the Nan\c
cay Radio Observatory, which is part of the Paris Observatory,
associated with the French Centre National de la Recherche
Scientifique. The Nan\c cay Observatory also gratefully acknowledges
the financial support of the Region Centre. We thank an anonymous
referee for comments which improved the paper.

\end{acknowledgements}

\end{document}